\documentstyle[aps,preprint,tighten,epsf]{revtex}
\def\myfigure#1#2{{\leftskip=0.10753\textwidth \rightskip\leftskip\small
\begin{figure}\baselineskip=14pt plus 2pt minus 1pt
\centerline{#1}\nobreak\smallskip\nobreak #2\end{figure}}}
\global\firstfigfalse
\begin{document}
\preprint{imsc/97/08/33}
\draft
\title{Coherent states for the Hydrogen atom}
\author{Pushan Majumdar \thanks{e-mail:pushan@imsc.ernet.in}\and 
H.S.Sharatchandra \thanks{e-mail:sharat@imsc.ernet.in}} 
\address{Institute of Mathematical Sciences,C.I.T campus Taramani.
Madras 600-113} 
\maketitle
\begin{abstract}
We construct wave packets for the hydrogen atom labelled by the 
classical action-angle variables with the following properties.
i) The time evolution is exactly given by classical evolution of the 
angle variables. (The angle variable corresponding to the position on the 
orbit is now non-compact and we do not get exactly the same state after 
one period. However the gross features do not change. In particular 
the wave packet remains peaked around the labels.)
ii) Resolution of identity using this overcomplete set involves exactly 
the classical phase space measure.
iii) Semi-classical limit is related to Bohr-Sommerfield quantization.
iv) They are almost minimum uncertainty wave packets in position and 
momentum.
\end{abstract}
\section{Introduction}

Schrodinger \cite{Sch} attempted to construct wave packets for the hydrogen 
atom
that were related to classical orbits. Such a construction was easy for
the harmonic oscillator, and these are the well known coherent states
\cite{Sud}\cite{Ska}. The hydrogen atom, however, proved to be more 
difficult and the
question was not resolved at that time. The issue has become relevant
again in connection with the Rydberg atoms \cite{Ryd} in microwave cavities.
Various considerations have led to different proposals \cite{H} for the 
coherent 
states of the hydrogen atom. Some used the dynamical groups SO(4) or 
SO(4,2). However, a state of the class did not go into a state of the 
same class under time evolution.
Klauder \cite{Kla} has constructed coherent states with the property that 
under time evolution these remain coherent states. 
Recently one of us \cite{Sharat} constructed a set of coherent states 
for the
anharmonic oscillator which was unique when precise connection to the
classical phase space and dynamics was demanded.
 
We construct wave packets for the hydrogen atom labelled by classical 
phase space variables.
\begin{eqnarray}
| R,\alpha,\beta,\gamma,\delta,\theta\rangle &=& \sum_{n=1}^{\infty} 
\;\sum_{j,m_{1},m_{2},l,m}
e^{-\frac{R}{2\hbar}}\frac{(R/\hbar)^{\frac{n-1}{2}}}{\sqrt{(n-1)!}} 
e^{-i\alpha m_{1}} e^{-i\gamma m_{2}} 
e^{i(\frac{R^{3}\theta }{2 n^{2} h^{3}})} \nonumber \\
&&\hspace{2cm} \times\frac{(2j)!}{\sqrt{(j+m_{1})!(j-m_{1})!
(j+m_{2})!(j-m_{1})!}} \nonumber \\
&&\hspace{1cm} \times sin(\frac{\beta}{2} )^{j-m_{1}} 
cos(\frac{\beta}{2})^{j+m_{1}}
sin(\frac{\delta}{2} )^{j-m_{2}} cos(\frac{\delta}{2})^{j+m_{2}}
C^{lm}_{jm_{1}jm_{2}}| n, l,m\rangle
\end{eqnarray}
The definition of these variables and their relation to the classical 
orbit is explained later.
The angle variable $\theta$ now has the range $(-\infty , \infty)$. Note 
that only the bound state spectrum has been used. Wave 
packets built out of scattering states with similar properties can also 
be constructed using our techniques, but will not be considered here. We 
have the resolution of identity in the subspace of the hilbert space 
spanned by the bound states,

\begin{eqnarray} {\bf 1}_{B.S.}&=&\frac{1}{h^{3}}\int_{0}^{\infty}\!dR
[\int ]\!d\theta \int_{-1}^{1}\!d((j+1/2)\hbar\, cos\,\beta)
\int_{0}^{2\pi}d\alpha \int_{-1}^{1}\!d((j+1/2)\hbar\, cos\,\delta)
\nonumber \\ &&\hspace{2in}\times \int_{0}^{2\pi} d\gamma | 
R,\alpha,\beta,\gamma,\delta,\theta\rangle
\langle R,\alpha,\beta,\gamma,\delta,\theta | 
\end{eqnarray} where
\begin{equation} [ \int ]d\theta \equiv \lim_{N\rightarrow \infty} 
\frac{1}{N}
\int_{-N\pi}^{N\pi}d\theta. \end{equation} This corresponds to averaging
over an infinite number of classical orbits. The measure is exactly the
classical phase space measure invariant under canonical transformations.
Under time evolution, 
\begin{equation} 
|R,\alpha,\beta,\gamma,\delta,\theta\rangle\stackrel{t}{\longrightarrow} 
|R,\alpha,\beta,\gamma,\delta,\theta+\omega (R) t\rangle
\end{equation}
 The wave packets peak around the point in the classical phase 
space represented by the labels. The expectation values of position and 
momenta do not exactly correspond to the labels and the wave packets are 
not of minimal uncertainty, in contrast to the harmonic oscillator 
coherent states. But these features of the latter are present in the 
semi-classical limit.

\section{Correspondence to the Kepler Problem}

The bound state Kepler problem is conveniently described \cite{Gold} by 
three action 
variables $M,L,R$ and their corresponding angle variables 
$\omega_{1},\omega_{2},\omega_{3}$. (These 
are related to the variables in Ref \cite{Gold} through $J_{3}=R, 
J_{2}=2\pi L, J_{3}=2\pi M, \omega_{i}=2\pi w_{i}, i=1,2,3$.)  
 The hamiltonian ${\cal H}$ involves only $R$.
\begin{equation}
{\cal H}=-\frac{2\pi^{2} me^{4}}{R^{2}}
\end{equation}
$L$ is the magnitude of the total angular momentum, 
and $M$ is the z component of the angular momentum. The angle 
variables $\omega_{1}$ and $\omega_{2}$ are also constants of motion in 
this problem because their corresponding frequencies are zero. Only 
$\omega_{3}$ changes in time as 
\begin{equation}
\omega_{3}(t)=\frac{2\pi}{T(R)}t\hspace{.5cm} {\text{with}} \hspace{.5cm}
\frac{1}{T(R)}=\frac{4\pi^{2}me^{4}}{R^{3}} 
\end{equation}
where T(R) is the time period of the orbit.

It has been observed in \cite{Kla}\cite{Sharat} that the time evolution in 
equation (4) is possible only if 
the angle variable $\omega_{3}$ is extended to the covering space; 
$\omega_{3}\in
(-\infty,\infty).$ This is because the energy levels are incommensurate. 
After one period the wave packet is not reproduced, though grossly it has 
the same features. This uniquely fixes the dependence on $\omega_{3}$.
\begin{equation}
|R,L,M,\omega_{1},\omega_{2},\omega_{3}\rangle = 
\sum_{nlm}C_{nlm}(R,L,M,\omega_{1},\omega_{2},)\, 
e^{\frac{-iE_{n}T(R)\omega_{3}}{2\pi\hbar}} |nlm\rangle
\end{equation}
Under rotation we require these wave packets to go into one another as 
these labels do.
\begin{equation}
|R,L,M,\omega_{1},\omega_{2},\omega_{3}\rangle\stackrel{\cal R}
{\longrightarrow} |R,L,M({\cal R}),\omega_{1}({\cal R}),\omega_{2}
({\cal R}),\omega_{3}\rangle 
\end{equation}
($R,L$ and $\omega_{3}$ do not change under rotation of axes.)

$\omega_{1}$ is the angle between the y axis and the line of nodes. 
(i.e. the 
line of intersection between the orbital plane and the x-y plane.)
 $\omega_{2}$ is the angle between the line of 
nodes and the major axis (fig 1). Also 
$\frac{M}{L}=cos\,\omega_{4}$, where $\omega_{4}$ is 
the inclination of the orbit. (i.e. the angle between the normal 
${\bf{n}}$ to 
the orbit and the z axis.) Thus under rotation of the orbit around 
the z axis (by angle $\psi_{1}$), $\omega_{1}$ increases by $\psi_{1}$ 
while $\omega_{2}$ and $\omega_{4}$ do not change. This uniquely requires 
the 
dependence on $\omega_{1}$ to involve $exp(-i\omega_{1}\hat{J_{z}})$ where 
$\hat{J_{z}}$ 
is the generator of rotations about the z axis. A rotation about the 
line of nodes by an angle $\psi_{4}$ increases $\omega_{4}$ by $\psi_{4}$ 
while keeping 
 $\omega_{1}$ and $\omega_{2}$ unchanged. This fixes the dependence on 
$\omega_{4}$ and 
$\omega_{1}$ so that it involves 
$exp(-i\omega_{1}\hat{J_{z}})exp(-i\,\omega_{4}\hat{J_{y}})$. This is 
because the rotation about line of nodes corresponds to $exp(-i 
\omega_{1}\hat{J_{z}})exp(-i 
\psi_{4}\hat{J_{y}})exp(\omega_{1}\hat{J_{z}})$. Finally a rotation 
about the normal ${\bf n}$ by angle $\psi_{4}$ increases 
$\omega_{2}$ 
by $\psi_{4}$ while keeping the other two angles constant. This rotation 
corresponds to 
\begin{displaymath}
exp(-i\omega_{1}\hat{J_{z}})\,exp(-i \,\omega_{4}\hat{ 
J_{y}})\,exp(-i\psi_{2} \hat{J_{z}})\,exp(i 
\,\omega_{4}\hat{J_{y}})\,exp(i\omega_{1}\hat{J_{z}}).
\end{displaymath}
Therefore the dependence on $\omega_{4},\,\omega_{1},\,\omega_{2}$ is 
required to be 
\begin{displaymath}
exp(-i\omega_{1}\hat{J_{z}})exp(-i \,\omega_{4}\hat{J_{y}})exp(-i
\omega_{2}\hat{J_{z}}).
\end{displaymath}
 Note that 
this rotation precisely corresponds to taking an orbit in the x-y plane 
with the major axis along the x direction into the orbit labelled by 
$(R,L,M,\omega_{1},\omega_{2},\omega_{3})$.

We may exploit the dynamical O(4) symmetry of the hydrogen atom to fix 
the dependence on $L$ also. In addition to the conserved vector {\bf 
J} related to rotational invariance, we have another conserved vector 
{\bf K} along the major axis (fig 2.) related to the Laplace Runge Lenz 
vector. We have ${\bf (J+K)}^{2}={\bf (J-K)}^{2}=R^{2}$, ${\bf 
J}^{2}=L^{2}$ and the eccentricity of the orbit is 
$e=\sqrt{1-\frac{L^{2}}{R^{2}}}$. The role of the vector {\bf K} 
is to deform the orbits by changing $L$. The O(4) symmetry 
corresponds to independent rotations of the vectors $\frac{\bf J+K}{2}$ and 
$\frac{\bf J-K}{2}$ in the 3-dimensional space.

Consider a circular orbit in the x-y plane. Now ${\bf K} = 0$ and ${\bf 
J\pm K}$ are in the z direction. Imagine a rotation of $\frac{\bf J+K}{2}$
about the 2-axis by an angle $\omega_{5}$ and an equal and opposite 
rotation of $\frac{\bf J-K}{2}$. This will give a non zero ${\bf K}$ of 
magnitude $R\, sin \,\omega_{5}$ along the x direction and ${\bf J}$ 
of magnitude $R\, cos \,\omega_{5}$ along the z direction. Thus the 
orbit has been deformed into an elliptic orbit in the x-y plane with 
$\frac{L}{R} = cos\,\omega_{5}$.

The above analysis shows the following. In order to have the right 
transformation properties of the classical variables 
$R,L,M,\omega_{1},\omega_{2},\omega_{3}$ under the full O(4) symmetry, the 
dependence on $R,L,\omega_{1},\omega_{2}$ has to be via 
\begin{displaymath}
e^{-i \omega_{1}\hat{J_{z}}}e^{-i \omega_{4}\hat{J_{y}}} 
e^{-i \omega_{2}\hat{J_{z}}}e^{-i \omega_{5}\frac{{\hat 
J_{y}}+{\hat K_{y}}}{2}} e^{i \omega_{5}\frac{{\hat J_{y}}-{\hat 
K_{y}}}{2}}
\end{displaymath}
where $cos\,\omega_{5}=\frac{L}{R}$ and 
$cos\,\omega_{4}=\frac{M}{L}$.
Classically this will rotate and deform a circular orbit in the x-y 
plane into the orbit with the labels 
$(R,L,M,\omega_{1},\omega_{2},\omega_{3})$ (without changing the size of 
the major axis). Quantum 
mechanically the former corresponds to the state $|n,n-1,n-1\rangle$.
Therefore we may expect the coherent state to have the form 
\begin{equation}
| R,L,M,\omega_{1},\omega_{2},\omega_{3}\rangle = \sum_{n}\,C_{n}(R)
e^{-i\omega_{1}\hat{J_{z}}}e^{-i \omega_{4}\hat{J_{y}}}
e^{-i\omega_{2}\hat{J_{z}}}
e^{-i \omega_{5}\frac{{\hat K_{y}}}{2}}
e^{i \frac{\omega_{3}R^{3}}{2 n^{2}h^{3}}}|n,n-1,n-1\rangle
\end{equation}
With a proper choice of $C_{n}(R)$ this will have the properties we 
require. However we find that it is much more natural and convenient to 
use a different set of action angle variables. Note the close relation 
to the angular momentum coherent states. Note also that the angle 
variables $\omega_{1},\omega_{2}$ are involved in rotation about the third 
axis 
whereas the angles $\omega_{4}$ and $\omega_{5}$ related to the action 
variables are involved in rotation about the one and two axes. This is 
a general feature as seen below.

\section{Coherent states for a precessing spin}

Consider a spinning object with spin quantum number j and gyromagnetic 
ratio $\mu$ in an 
external magnetic field B in the z direction. The hamiltonian is 
$\hat{{\cal H}}=\mu B \hat{J}_{z}$. Classically the spin will 
precess about the z 
axis with frequency $\mu B$. The action variable is $J_{z}$ which 
measures the inclination to the z axis and the angle 
variable $\theta \in (0,2\pi )$, is the azimuthal angle of the 
precessing spin. We now show that, by requiring classical time evolution, 
semi-classical limit and correct rotation property for the states 
$|J_{z},\theta\rangle$ labelled by the classical phase space of this 
system, we obtain uniquely the angular momentum coherent states \cite{Per}
. We have 
\begin{equation}
|J_{z},\theta \rangle = \sum_{m} C_{m}(J_{z}) 
e^{-\frac{i}{\hbar}\mu Bm\hbar\frac{\theta}{\mu B}}|j,m\rangle
\end{equation}
to reproduce the classical evolution, $|J_{z},\theta \rangle\stackrel{t}
{\longrightarrow} |J_{z},\theta +\mu Bt\rangle. $ Under rotation by 
angle $\psi$ about the x axis, $\omega$ goes to $\omega+\psi$ where 
$cos\,\omega=\frac{J_{z}}{J}$ and $J$ is the classical spin to be 
associated to the spin quantum number $j$. In order 
that $|J_{z},\theta \rangle$ have this property, we have to choose 
\begin{equation}
|J_{z},\theta \rangle=e^{-i \theta\hat{J_{z}}}e^{-i \omega\hat{J_{y}}}
|jj\rangle
\end{equation}
This is precisely the rotation that takes the z axis to the 
instantaneous axis of the classical spin. Correct semi-classical limit 
requires the choice $|jj\rangle$ as seen below. Note
that we have precisely got the angular momentum coherent state 
labelled by $\omega$ and $\theta$.
We now show that this has the right semi classical limit and resolution 
of identity
\begin{equation}
|J_{z},\theta \rangle=\sum_{m}d^{j}\!_{jm}(\omega)e^{-im\theta}
|jm\rangle
\end{equation}
where 
\begin{equation}
d^{j}\!_{jm}(\omega)=\sqrt{\frac{(2j)!}{(j+m)!(j-m)!}} sin(\frac{\omega}{2}
)^{j-m} cos(\frac{\omega}{2})^{j+m}
\end{equation}
For large $j$, $d^{j}\!_{jm}(\omega)$ peaks at $cos\,\omega =\frac{m}{j}$ 
i.e. the dominant contributions come from the states 
$m\hbar\approx J_{z}$.

As $J_{z}$ and $\theta$ are action angle variables, the phase space measure 
is $dJ_{z}\,d\theta$. Now
\begin{eqnarray}
\frac{1}{h}\int_{-J}^{J}dJ_{z}\int_{0}^{2\pi}d\theta\,|J_{z},\theta \rangle
\langle J_{z},\theta |&=&\sum_{m}\frac{J}{\hbar}\int_{-1}^{1}d(cos\,\omega)
\;d^{j}\!_{jm}(\omega)\; 
d^{j}\!_{jm}(\omega) |jm\rangle\langle jm| \\
&=& \frac{J}{\hbar}\frac{2}{2j+1}|jm\rangle\langle jm| \\ &=&{\bf 1}
\end{eqnarray}
with the identification $J=(j+1/2)\hbar$. (This means that we must 
associate classical $J=(j+1/2)\hbar$ to the spin quantum number $j$.)

Thus the angles $\omega$ and $\theta$ appearing in the angular momentum 
coherent state Eq. (12) can be interpreted as classical phase 
space 
variables for a precessing spin with $\theta$ as the angle variable 
and $(j+1/2)cos\,\omega$ as the corresponding action variable.

\section{Coherent states for the hydrogen atom}

In place of the conserved variables $L,M,\omega_{1},\omega_{2}$ we will use 
other variables suggested by the O(4) symmetry. We will use the two O(3) 
subgroups 
in O(4) generated by $(\frac{\hat{\bf J}\pm\hat{\bf K}}{2})$. We define 
\begin{eqnarray}
| R,\alpha,\beta,\gamma,\delta,\theta\rangle &=& \sum_{j}C_{j}(R)
e^{-i\alpha{(\frac{\hat{J}+\hat{K}}{2})_{z}}}
e^{-i\beta{(\frac{\hat{J}+\hat{K}}{2})_{y}}}
e^{-i\gamma{(\frac{\hat{J}-\hat{K}}{2})_{z}}}
e^{-i\delta{(\frac{\hat{J}-\hat{K}}{2})_{y}}} 
e^{i (\frac{R^{3}\theta}{2 n^{2} h^{3}})} |jj\rangle |jj\rangle
\end{eqnarray}

In place of quantum states $|n,l,m\rangle$ we are now using
$|jm_{1}\rangle |jm_{2}\rangle$ of $\frac{\hat{\bf J}+\hat{\bf K}}{2}$
and $\frac{\hat{\bf J}-\hat{\bf K}}{2}$ respectively. (The $j$ quantum
number is the same because $(\hat{\bf J}+\hat{\bf K})^{2}=(\hat{\bf
J}-\hat{\bf K})^{2}$). $j$ takes half integer values
$0,\frac{1}{2},1,\frac{3}{2}...$ We get the states $ |nlm\rangle$ by
going to the coupled basis 
\begin{equation} 
|jm_{1}\rangle |jm_{2}\rangle
= \sum_{lm} C^{lm}_{jm_{1}jm_{2}}|2j+1,l,m\rangle. 
\end{equation}

The new angles are related to the earlier angles as follows (ref fig.2).
The $\frac{\hat{\bf J}+\hat{\bf K}}{2}$ rotation rotates the classical
vector ${\bf J}+{\bf K}$ from the z axis to $(R\,sin\,\beta\,cos\,\alpha,
R\,sin\,\beta\,sin\,\alpha, R\,cos\,\beta)$ without affecting
$\frac{\hat{\bf J}-\hat{\bf K}}{2}$. Similarly the $\frac{\hat{\bf 
J}-\hat{\bf K}}{2}$
rotation rotates the classical vector ${\bf J}-{\bf K}$ from z axis to
$(R\,sin\,\delta\,cos\,\gamma,R\,sin\,\delta\,sin\,\gamma,
R\,cos\,\delta).$ Therefore the projection of {\bf J} on the z axis gives
$cos\,\psi_{4}=\frac{R}{|{\bf J}|}(cos\,\beta + cos\,\delta)$ where 
$|{\bf J}|=R\sqrt{2+2\,sin\,\beta\,sin\,\delta\, 
cos(\alpha - \gamma)\,+\,2\,cos\,\beta\,cos\,\delta}.$
 The line of nodes is along $\hat{z}\times\hat{J}$ and therefore has
the direction cosines $(sin\,\delta\,sin\,\gamma +
sin\,\beta\,sin\,\alpha,-sin\,\beta\,cos\,\alpha-sin\,\delta\,cos\,\gamma)$.
Therefore $cos\,\omega=\frac{1}{|{\bf ON}|}(sin\,\delta\,sin\,\gamma +
sin\,\beta\,sin\,\gamma)$ with $|{\bf 
ON}|=R\sqrt{sin^{2}\beta+sin^{2}\gamma + 
2\,sin\,\beta\,sin\,\gamma\,cos(\alpha - \delta)}$ . $\Omega $ is 
obtained by taking the 
component of {\bf K} along the line of nodes and therefore $cos\:\Omega=
\frac{1}{|{\bf ON}||{\bf K}|}
(sin^{2}\beta\,cos\,2\alpha-sin^{2}\gamma\,cos\,2\delta)$ and $|{\bf 
K}|=R\sqrt{2-2\,sin\,\beta\,sin\,\delta\,
cos(\alpha - \gamma)\,-\,2\,cos\,\beta\,cos\,\delta}.$
The orbit is simply obtained from the vectors {\bf J}+{\bf K} and {\bf
J}-{\bf K} because it is perpendicular to {\bf J} and has the major axis
along the direction ${\bf K}$ with magnitude $(\frac{R}{2\pi h})^{2}a$ 
where $a$ is the Bohr radius. Also the eccentricity is 
given by $e=\sqrt{1-\frac{{\bf J}^{2}}{({\bf J}\pm {\bf K})^{2}}}.$

The classical phase space measure in the new variables is
\begin{displaymath}
dR\,d\theta\,d((j+1/2)\hbar\,cos\,\beta)\,d\alpha\,d((j+1/2)\hbar\,cos\,\delta)
\,d\gamma
\end{displaymath}
For large $J$, the state $| R,\alpha,\beta,\gamma,\delta, \theta\rangle$
gets dominant contribution from $m_{1}=(j+1/2)\,cos\,\beta$ and
$m_{2}=(j+1/2)\,cos\,\delta $. This is exactly as wanted by Bohr
quantization of the action angle pairs because $cos\:\beta=
\frac{(J+K)_{z}}{R}$ and $cos\,\delta= \frac{(J-K)_{z}}{R}$.
Therefore we only have to fix $C_{j}(R)$ by requiring the correct 
semi-classical limit and resolution of identity.
We want $C_{j}(R)$ to peak at $R=(2j+1)\hbar$ as Bohr quantization 
gives $R=n\hbar$. Also to get a resolution of identity we require 
\begin{equation}
\frac{1}{\hbar}\int_{0}^{\infty}dR\,|C_{j}(R)|^{2}=1
\end{equation}
for all $j$. For normalization we require $\sum_{j}|C_{j}(R)|^{2}=1$ where 
$j=0,\frac{1}{2},1,\frac{3}{2}..$. All these requirements are met by 
\begin{equation}
C_{j}(R)=e^{-\frac{R}{2\hbar}}\frac{(R/\hbar)^{j}}{\sqrt{(2j)!}}
\end{equation}
Thus we get the coherent state as in equation no.(1).

\section{Wave packet properties}

In case of the harmonic oscillator coherent states $|z\rangle$ the 
expectation values of the position and momentum operators are directly 
given by the real and imaginary parts of the label $z$. Also they are 
minimal uncertainty states. For our coherent states, these properties 
are not valid exactly, but are valid asymptotically in the 
semi-classical region \cite{Sharat}. This is a consequence of the 
semi-classical 
limit of our coherent states where the correspondence principle may be 
applied. Consider the expectation value of an operator 
$\hat{O}(\hat{p,q})$ in a coherent state. For large values of $R,L,$ and 
$M$ (in units of $\hbar$), the coherent state is dominated by the states 
$|nlm\rangle$ with $n\approx R/\hbar$, $l\approx L/\hbar$ and $m\approx 
M/\hbar$. Now, the correspondence principle relates the expectation 
value of $\hat{O}$ to the value of the corresponding classical variable 
$O(p,q)$ for the corresponding classical orbit. Thus asymptotically, our 
coherent states are wave packets peaked around position, momenta etc. 
corresponding to the action angle variables labelling it. Also, 
asymptotically they would be minimum uncertainty states. More detailed 
consideration of these properties for small values of the action 
variables will be considered elsewhere.

\section{Conclusion}

We have constructed wave packets for the hydrogen atom, labelled by 
points of the classical phase space which follow classical orbits very 
closely. They have the correct semiclassical limit corresponding to Bohr 
quantization. In addition, they have the desirable property that the 
resolution of identity involves exactly the classical phase space measure.
As a consequence of incommensurate energy levels, our wave packets do 
not return to the original state after one period, but the overall 
features do not change. One may interpret this as follows : the wave 
packet has (an infinite number of) internal degrees of freedom, which 
may not return to the original state after a period.

\section{Acknowledgements}

One of us (H.S.Sharatchandra) thanks Professor K.H.Mariwalla for helpful 
discussions and pointing out references.

\newpage
\myfigure{\epsfysize 3.375in\epsfbox{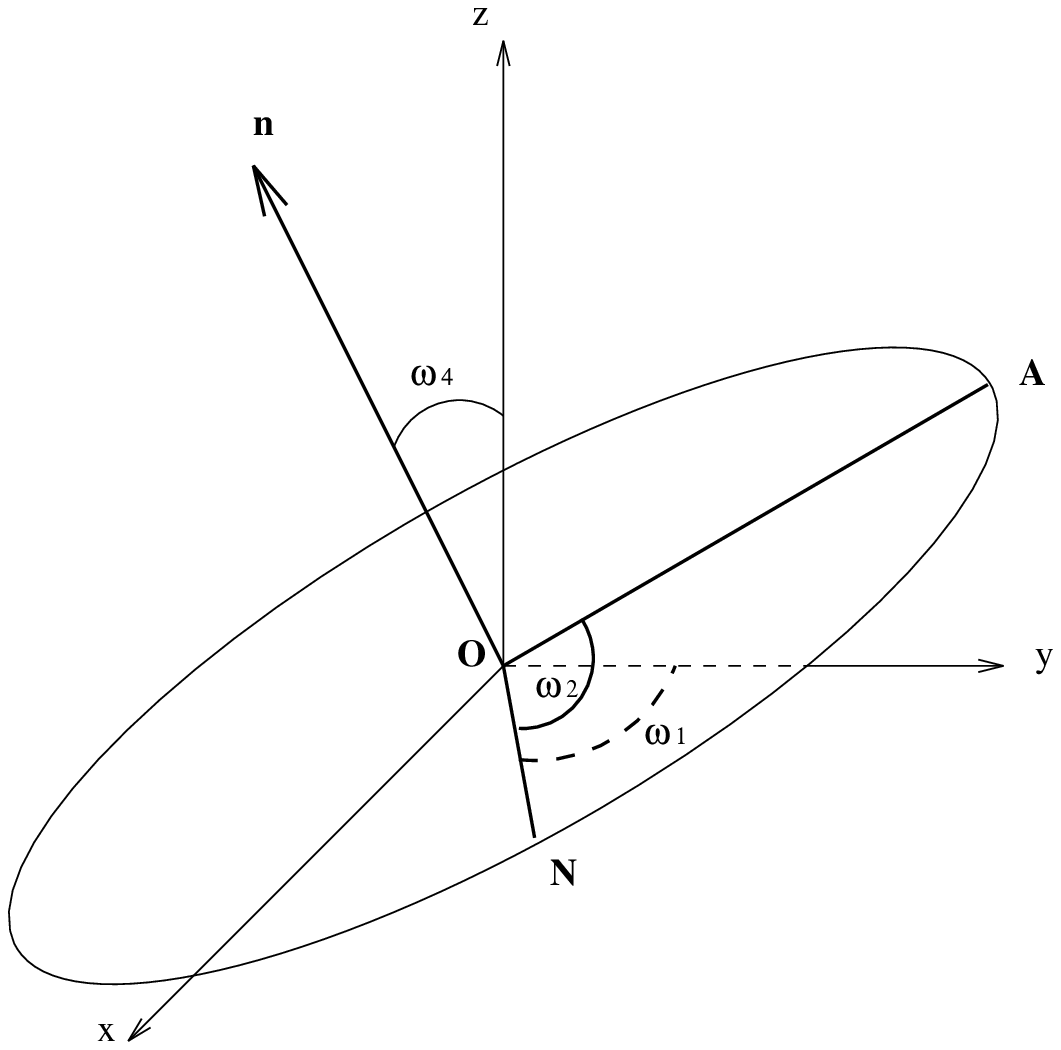}}{
\begin{center}fig. 1. The classical elliptic orbit. 
{\bf ON}: Line of Nodes. {\bf OA}: Major axis.\end{center}}

\myfigure{\epsfysize 3.375in\epsfbox{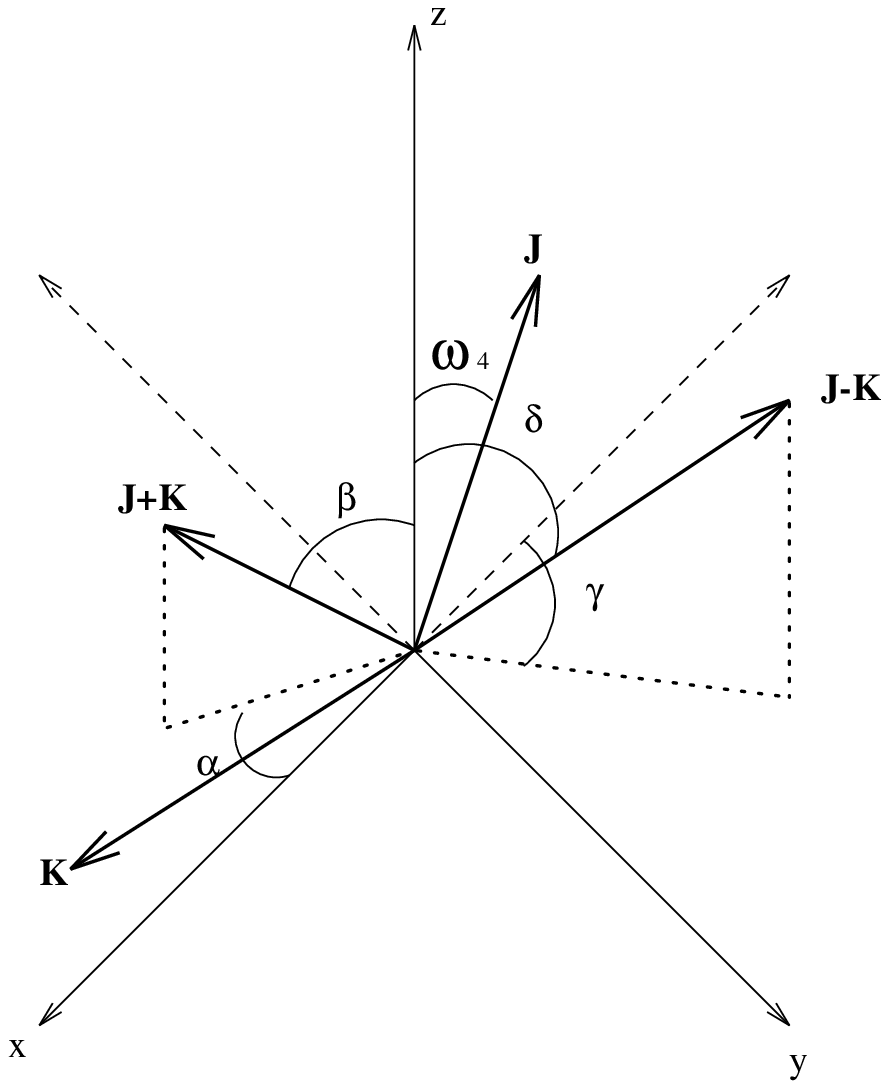}}{\begin{center}fig. 2. The
vectors {\bf J} and {\bf K} and the angles associated with
them.\end{center}}

\end{document}